\documentclass[preprint,showpacs,amssymb,amsmath,superscriptaddress]{revtex4-1}
\usepackage{graphicx,color}% Include figure files
\usepackage{dcolumn}% Align table columns on decimal point

\newcommand{\micron}{\,\mu\textrm{m}}
\newcommand{\nm}{\,\textrm{nm}}

\begin{document}

\title{Experimental observation of optical Weyl points}

%% Notice placement of commas and superscripts and use of &
%% in the author list

\author{Jiho Noh*}
\affiliation{Department of Physics, The Pennsylvania State University, University Park, PA 16802, USA}
\author{Sheng Huang*}
\affiliation{Department of Electrical and Computer Engineering, University of Pittsburgh, Pittsburgh, Pennsylvania 15261, USA}
\author{Daniel Leykam*}
\affiliation{School of Physical and Mathematical Sciences, Nanyang Technological University, Singapore 637371, Singapore}
\author{Y.~D.~Chong}
\affiliation{School of Physical and Mathematical Sciences, Nanyang Technological University, Singapore 637371, Singapore}
\affiliation{Centre for Disruptive Photonic Technologies, Nanyang Technological University, Singapore 637371, Singapore}
\author{Kevin Chen}
\affiliation{Department of Electrical and Computer Engineering, University of Pittsburgh, Pittsburgh, Pennsylvania 15261, USA}
\author{Mikael C. Rechtsman}
\affiliation{Department of Physics, The Pennsylvania State University, University Park, PA 16802, USA}

\date{\today}

\begin{abstract}
Weyl fermions are hypothetical two-component massless relativistic particles in three-dimensional (3D) space, proposed by Hermann Weyl in 1929.  Their band-crossing points, called ``Weyl points'', carry a topological charge and are therefore highly robust.  There has been much excitement over recent observations of Weyl points in microwave photonic crystals and the semimetal TaAs.  Here, we report on the first experimental observation of Weyl points of light at optical frequencies.  These are also the first observations of ``type-II'' Weyl points for photons, which have strictly positive group velocity along one spatial direction.  We use a 3D structure consisting of laser-written waveguides, and show the presence of type-II Weyl points by (1) observing conical diffraction along one axis when the frequency is tuned to the Weyl point; and (2) observing the associated Fermi arc surface states.  The realization of Weyl points at optical frequencies allow these novel electromagnetic modes to be further explored in the context of linear, nonlinear, and quantum optics.     
\end{abstract}
%\pacs{42.70.Qs, 03.65.Vf, 73.20.At}
\maketitle

The observation of Weyl points, in microwave photonics~\cite{lu2015} and condensed matter physics~\cite{xu2015,lv2015}, attracted a great deal of attention because they constitute the simplest possible topologically-nontrivial band structure in 3D.  Their topological protection gives rise to the unique feature of robust ``Fermi arc'' surface states~\cite{fermi_arcs,fermi_arcs_2,burkov2011,potter2014quantum}, and they can be associated with many interesting phenomena including chiral anomalies~\cite{chiral_1,chiral_2}, unconventional superconductivity~\cite{cho2012superconductivity}, and large-volume single-mode lasing~\cite{bravoabad2012}. Weyl points take two distinct forms, type-I and type-II, which have point-like and conical Fermi surfaces respectively~\cite{soluyanov2015}.  In photonics, type-I Weyl points were predicted~\cite{lu2013} and subsequently observed~\cite{lu2015} in macroscopic photonic crystals at microwave frequencies.  There have been significant efforts, both theoretical~\cite{lu2013, Wang2016Weyl, bravo2015weyl, gao2015plasmon, xiao2015synthetic, yang2016acoustic, PhysRevLett.117.057401} and experimental~\cite{chen2015experimental, atwater, peng2016gyroid}, to realize Weyl points at the technologically important optical frequency regime; this is challenging, however, due to the need for 3D fabrication of intricate photonic crystal structures.  Photonic type-II Weyl points have also been theoretically proposed \cite{xiao2016hyperbolic}, though not experimentally observed.
 
Here we observe photonic type-II Weyl points, at optical frequencies, in a 3D photonic crystal structure consisting of evanescently-coupled waveguides. Type-II Weyl points are distinguished by the strongly anisotropic dispersion depicted schematically in Fig. 1, with a conical isofrequency surface in a particular direction. The waveguide array that we employ is fabricated using femtosecond direct laser writing~\cite{szameit_review}.  The waveguides form an array aligned along a special spatial axis $z$, as shown in Fig.~2(a).  In previous works, optical diffraction through such an array has been conceptualized in terms of a reduced-dimensionality system~\cite{yariv1984optical}: in the paraxial limit, the envelope function of the linearly-polarized electric field, $\psi$, is found to obey a (2+1)D Schr\"odinger wave equation
\begin{equation}
i\partial_z \psi = -\frac{1}{2k_0}\nabla_\perp^2 \psi - \frac{k_0 \delta n(x,y,z)}{n_0}\psi, \label{eq:se}
\end{equation}
with the $z$ direction acting as ``time'' and the two transverse dimensions ($x$ and $y$) acting as ``space''.  Here, $k_0=2\pi n_0/\lambda$ is the wavenumber in the ambient medium (Corning Eagle XG borosilicate glass, index $n_0=1.5078$), $\lambda$ is the wavelength of the input light, and $\delta n$ describes the refractive index difference produced by the waveguides.  We take $\delta n$ to be periodic in the $z$ direction, as well as in $x$ and $y$, so that Eq.~(\ref{eq:se}) corresponds to a Floquet lattice (i.e., a periodically-driven crystal).  The operating frequency, $\omega = 2\pi c/\lambda$, enters into the Hamiltonian of the (2+1)D system as a tunable parameter.

But we can also regard the waveguide array as a 3D photonic crystal, with a photonic band structure that describes the frequency $\omega$ as a function of the Bloch wavevector $(k_x,k_y,k_z)$.  We shall argue that this 3D band structure can exhibit type-II Weyl points.  This occurs if, and only if, the (2+1)D Floquet band structure of Eq.~(\ref{eq:se}) reduces to a 2D Dirac equation in which the Dirac mass can be tuned through zero by varying $\omega$.  At a specific value of $\omega$, the 3D band structure possesses a spectral degeneracy in the form of a Weyl point; correspondingly, the (2+1)D system at that parametric choice of $\omega$ possesses a spectral degeneracy in the form of a massless 2D Dirac equation. The Dirac dispersion that arises in the isofrequency surface of the Weyl dispersion is depicted in Fig. 1(b).

Our 3D structure, depicted in Fig.~2(a), consists of a bipartite square lattice composed of two helical waveguides in each unit cell, both having clockwise helicity, radius $R$, and period $Z$ in the $z$ direction.  The helical waveguide modulation breaks the inversion symmetry of the structure---a necessary condition for the realization of Weyl points in time-reversal-invariant systems such as this one.  Unlike the helical waveguide array used to realize a photonic topological insulator in Ref.~\cite{rechtsman2013}, adjacent waveguides are out of phase by $\pi$ (i.e., a half-cycle).  Thus, they do not evolve in synchrony, but come closer and farther from one another over the course of a cycle in $z$ (this structure is described in detail in Ref.~\cite{leykam2016}). We fabricated the waveguide array in borosilicate glass using a Titanium:Sapphire laser and amplifier system (Coherent:RegA 9000) with pulse duration 270fs, repetition rate 250kHz, and pulse energy 950nJ. The laser writing beam was sent through a beam shaping cylindrical telescope to control the shape and size of the focal volume. The spatial beam shaping was critical to produce symmetrical waveguide profiles, yielding a small coupling strength ratio of less than 1.1:1 (Cv:Ch, obtained by fitting linear array mode patterns along two directions). The beam was then focused inside the glass chip using an 80X, aberration-corrected microscope objective (NA = 0.75).  A high-precision three-axis Aerotech motion stage (model ABL20020) is used to translate the sample during fabrication. Experiments are performed by butt-coupling a single mode optical fiber to waveguides at the input facet of the chip, which subsequently couples to the waveguide array. The input light is supplied by a tunable mid-infrared diode laser (Agilent 8164B), which can be tuned through the 1450\,\nm--1650\,\nm wavelength range. After a total propagation distance of $4\,\text{cm}$ within the array, the light output from the waveguide array is observed using a 0.2\,NA microscope objective lens and a near-infrared InGaAs camera (ICI systems).  A microscope image of the output facet of the structure is shown in Fig.~2(b).

We can compute the (2+1)D Floquet band structures, in the paraxial limit, using numerical techniques similar to those employed for Floquet topological insulators~\cite{oka2009photovoltaic, Demler10, lindner2011floquet, rechtsman2013, leykam2016}.  The results are shown in Figures~2(d)-(f) for helix radius $R=4\,\micron$ and period $Z=1\,\text{cm}$, using three different values of the lattice constant ($a=29\micron$, $a=27\micron$, and $a=25\micron$) and operating wavelength ($\lambda = 1450\nm$, $\lambda = 1525\nm$, and $\lambda = 1600\nm$).

We have emphasized that $\lambda$, or equivalently the frequency $\omega$, enters into Eq.~(\ref{eq:se}) as a tunable parameter.  For each $\omega$, the Floquet bands form isofrequency contours of the 3D photonic band structure, as a function of $(k_x,k_y,k_z)$. This correspondence can be established rigorously, by showing that a 2D Dirac equation, with Floquet quasienergy $\beta$ (the eigenvalue of the Floquet Hamiltonian) and mass depending parametrically on $\omega$, can be mapped onto a 3D Weyl Hamiltonian with eigenvalue $\omega$ and depending parametrically on the propagation constant $k_z = \beta + k_0$.  Details of this mapping are given in the Supplementary Information.

In Fig.~2(d), we see that for $a=29\,\micron$ and $\lambda = 1450\,\nm$, the Floquet band structure has a topologically trivial band gap at quasienergy $\beta = \pi/Z$.  As we decrease $a$ and/or increase $\lambda$, the band gap decreases in size and eventually closes at Bloch wavevector ${\bf q}=\Gamma=(0,0)$.  This shows the presence of a conical isofrequency surface (Fig.~2(e)), which is the manifestation of the Weyl point in the 3D band structure (see Supplementary Information Section 1).  For still smaller $a$ and larger $\lambda$, the bulk band gap reopens after having undergone a topological transition; this gap is topological, and thus associated with gapless chiral edge states along the lattice edge (Fig.~2(f)).  The topological transition serves as a key observable indicator for the presence of the Weyl point with quantized nonzero Berry flux. In terms of the 2D Dirac equation, it signifies that the Dirac mass $m$ (gap size) changes sign by crossing zero; small perturbations cannot destroy the band crossing point, only shift it. Fig.~2(c) shows the calculated topological phase diagram as a function of $a$ and $\lambda$, confirming the robustness of the band crossing.  In particular, increasing $a$ shifts the crossing to longer wavelengths---but the transition necessarily occurs at a particular wavelength, $\lambda$.  This robustness reflects the topological protection of the Weyl point in the 3D band structure.

We prove the existence of the Weyl point in two distinct experiments: (1) by observing conical diffraction associated with the conical isofrequency surface of a type II Weyl point~\cite{soluyanov2015}; and (2) by observing the Fermi arc surface states that emerge from the Weyl point.  

Since the isofrequency cut of the type-II Weyl point is a 2D Dirac cone (Fig.~2(d)), every in-plane Bloch wave has the same transverse group velocity (the effective ``speed of light'' of the relativistic massless particle), regardless of the Bloch wavevector $\boldsymbol{k}$. Therefore, an initial wavepacket, which can be decomposed into a superposition of Bloch waves, evolves into a ring whose radius expands at the band speed.  In the context of paraxial optics, this corresponds to the phenomenon of \textit{conical diffraction}, marked by a fixed angle of diffraction relative to the axis of the waveguide array~\cite{berry_conical,peleg2007}. The presence of conical diffraction at a \emph{single} probe wavelength distinguishes our Weyl point from gapped phases or Dirac line nodes~\cite{line_node}.  In the latter class of systems, which include honeycomb photonic lattices~\cite{peleg2007}, conical diffraction is observable over a continuous range of probe wavelengths.

To demonstrate conical diffraction, we excite the system by injecting light into a single waveguide at the input facet, which then couples through a waveguide splitter to a pair of neighboring waveguides (within one unit cell at the center of the lattice) with equal intensity and phase.  The in-coupling region occupies the first $1\,\text{cm}$ of the chip.  The resulting diffraction patterns, imaged at the end of the chip, are shown in Fig.~3(a-c).  For lattice constants $a=29\micron$ and $a=25\micron$, for which the Floquet band structure is in a conventional insulator and topological insulator phase respectively, we observe a filled-in disc-like diffraction pattern, which is characteristic of band edge modes.  When the lattice constant is tuned to $a=27\micron$, we observe a clear ring-like conical diffraction pattern, shown in Fig.~3(b).  Deviations from a perfect ring occur because the lattice is discrete, which makes it impossible for the light to reside on a perfectly circular and zero-width ring.  Additionally, the diffraction pattern is distorted by the square symmetry of the lattice, which induces a saddle point in the quasienergy at the Brillouin zone edge.  Fig.~3(d-f) show full-wave beam-propagation simulations, which agree strongly with the experimental results of Fig.~3(a-c).

To quantify the observation of conical diffraction, we define a dimensionless parameter that measures the degree to which the diffraction pattern is conical:
\begin{equation}
C = \frac{\displaystyle \int d{\bf r} \; r^2 \, \left|\psi({\bf r})\right|^2}{\displaystyle \left(\int d{\bf r}\; r\, \left|\psi({\bf r})\right|^2\right)^2},
\end{equation}    
where ${\bf r}=(x,y)$ and $r=\sqrt{x^2+y^2}$ is the distance from the origin (the origin is defined to be the center point between the two excited waveguides).  The quantity $C$ measures how ``ring-like'' a wavefunction is, and is reminiscent of the inertial moment of a rotating body in a mechanical system (but is dimensionless in this case).  It takes the value 1 for an infinitely thin ring, and is larger than 1 for all other patterns.  In Fig.~3(g-i) we plot $C$ against wavelength for the three previous lattice constants, namely $a=29,27,25\micron$.  
The monotonic behavior of Figs. 3(g) and 3(i) show that for lattice constants $a=29\mu m, 25\mu m$ the Weyl point does not occur within the wavelength range of the tunable laser ($1450$--$1650\nm$), or is very near the boundary (i.e., the gap is small but does reach zero).  However, the minimum in $C$ observed in Fig.~2(h) corresponds to conical diffraction (see Figs. 3(b) and 3(e) for experimental and numerically computed conical diffraction patterns).  The conical diffraction associated with conical isofrequency surface provides direct evidence of the existence of the type-II Weyl point.  

As stated earlier, the Weyl point in the 3D band structure corresponds to a topological transition in the (2+1)D Floquet band structure.  To study this, we use the fact that the topologically trivial phase of the lattice (large $a$) should have no edge states, whereas the topologically nontrivial phase (small $a$) has chiral unidirectional edge states.  We demonstrate this by injecting light via a single waveguide along the top edge of the lattice, and observing the evolution of the wavepacket.  The first row of Fig.~4 shows the output wavepacket at wavelength $1550\nm$, with decreasing lattice constant from left to right.  Figure 4(a,b) show the result deep in the topologically trivial regime ($a=29, 28\,\mu$m): light that is input onto the edge spreads into the bulk due to the lack of an edge state. However, in the topological region ($a=26, 25\,\mu$m, Fig.~4(d,e)), most of the light stays confined to the edge and proceeds in a clockwise sense, an indicator of the chiral edge state.  In fact, the topological phase observed here is anomalous~\cite{rudner2013anomalous, gao2016probing, maczewsky2016observation, mukherjee2016experimental}, in the sense that despite the presence of a chiral edge state, the single band in the 2D Floquet band structure has Chern number zero.  Its associated topological invariant is described in detail in ~\cite{rudner2013anomalous}.   Full-wave beam-propagation simulations that correspond to the cases shown in Fig.~4(a-e) are shown in Figs. 3(f-j).  Corresponding edge band structures are shown in Figs. 4(k-o), indicating the transition from trivial to topological regimes.  Clearly, the band structures show that states confined to opposite edges (marked in red and blue) emerge after the topological transition at the Dirac point.  This acts as a further piece of evidence of the existence of the Weyl point at the transition between the two phases.
Note that we can also see the topological transition by varying $\lambda$ instead of $a$, as shown by the phase diagram of Fig.~2(c); we have chosen to vary the latter to make the trend more obvious, since the Dirac gap is strongly dependent on $a$.

The edge states observed here also constitute the ``Fermi arc'' surface states that are predicted to emerge from Weyl points~\cite{fermi_arcs,fermi_arcs_2,burkov2011,potter2014quantum}.  Indeed, the dispersion diagrams shown in Figs.~4(m-o) for the edge states in the transverse $(x,y)$-plane are equifrequency contours of Fermi arc states occupying the surface of the 3D structure.  Their observation at the Weyl point can be taken as further experimental evidence of the presence of the Weyl point.

Finally, let us discuss the robustness of our method for identifying Weyl points in 3D based on the (2+1)D Floquet band structure of Eq.~(\ref{eq:se}).  For the (2+1)D system, we have taken the paraxial limit, which means assuming that the two propagation directions ($\pm z$) and polarization degrees of freedom ($x,y$) are decoupled.  However, Weyl points are protected because they carry a quantized topological charge (chirality or Berry flux); they can only be destroyed by annihilating with a Weyl point of opposite charge. Since our system has time-reversal symmetry, reversing the propagation direction yields another Weyl point, with the opposite sign of $k_z$, but identical charge and frequency~\cite{lu2015}. Moreover, the Weyl points in the orthogonal polarization not only have identical charges, but they are also split along the $k_z$ axis as a result of the structural birefringence of the individual waveguides, which have elliptical cross-sections (see Fig. 2(b)). This structural birefringence is quantified as the difference in the waveguide propagation constant for the two polarizations.  Using a full diagonalization of Maxwell's equations, we calculated this to be $2.0\times 10^{-6}\,\micron^{-1}$. Hence, these additional degrees of freedom cannot destroy our Weyl point.

The above reasoning implies that partner Weyl points of opposite charge exist elsewhere in the 3D band structure.  To find them, we can use the fact that they must serve as the other (long-wavelength) termination points of the Fermi arc surface states found in the nontrivial phase of Fig.~2(c). Numerical simulations reveal that this occurs outside the range of our laser, but is accessible using shorter lattice periods. For example, in a lattice of period $a=22\,\micron$ the band gap closes and the edge modes terminate at $\lambda = 1650\,\nm$. Our Weyl point pair is separated in frequency, which is a generic property of type-II Weyl points~\cite{soluyanov2015}, but not realized in previous Weyl photonic crystal designs~\cite{lu2013}. The lack of a degenerate partner Weyl point of opposite charge has some interesting applications; for instance, one could couple a localized emitter to the chiral Weyl dispersion, and exploit the fact that the Dirac point in the Floquet Hamiltonian is {\it unpaired}, implying the lack of coherent backscattering and Anderson localization in the presence of disorder~\cite{AndersonWhitePaint, LagendijkParaxial, Wiersma1997, ChabanovGenack2000, Maret2006, SchwartzMoti, SilberbergAndLoc,leykam2016}.

In conclusion, we have made the first direct experimental observations of a Weyl point at optical frequencies, which is a type-II Weyl point.  We observed conical diffraction occurring at a single frequency (corresponding to a topological transition of a 2D Dirac equation), as well as surface states corresponding to Fermi arcs emerging from the Weyl point.  The observation of Weyl points in optics can lead to a range of novel phenomena arising from the interplay of the Weyl dispersion and nonlinear~\cite{nonlinear}, non-Hermitian~\cite{BoExceptionalPoints}, and quantum optics~\cite{peruzzo2010quantum, quantumtopo}. \\

%% Natural open questions include: what will be the effect of non-Hermiticity (loss/gain/lasing) near the Weyl point?  Are states in the vicinity of the Weyl point modulationally stable in the presence of Kerr nonlinearity, or other optical nonlinearities?  Can Weyl points be gapped by aperiodic strain, and if so can solitons reside in those band gaps?  Will the Weyl point have a non-trivial effect on photon pair generation and the resulting photon correlations?  The introduction of Weyl points into the optical domain allows these and countless other new directions of exploration.    

\noindent {\bf Acknowledgements} M.C.R. acknowledges the National Science Foundation under award number ECCS-1509546, the Penn State MRSEC, Center for Nanoscale Science, under award number NSF DMR-1420620, and the Alfred P. Sloan Foundation under fellowship number FG-2016-6418.  D.L. and C.Y.D. acknowledge
support by the Singapore National Research Foundation under grant
No.~NRFF2012-02, by the Singapore MOE Academic Research Fund Tier 2
Grant No. MOE2015-T2-2-008, and by the Singapore MOE Academic Research
Fund Tier 3 grant MOE2011-T3-1-005.  K.C. acknowledges the National Science Foundation under award number ECCS-159199.  

\newpage

\begin{figure}
\includegraphics[width=6in]{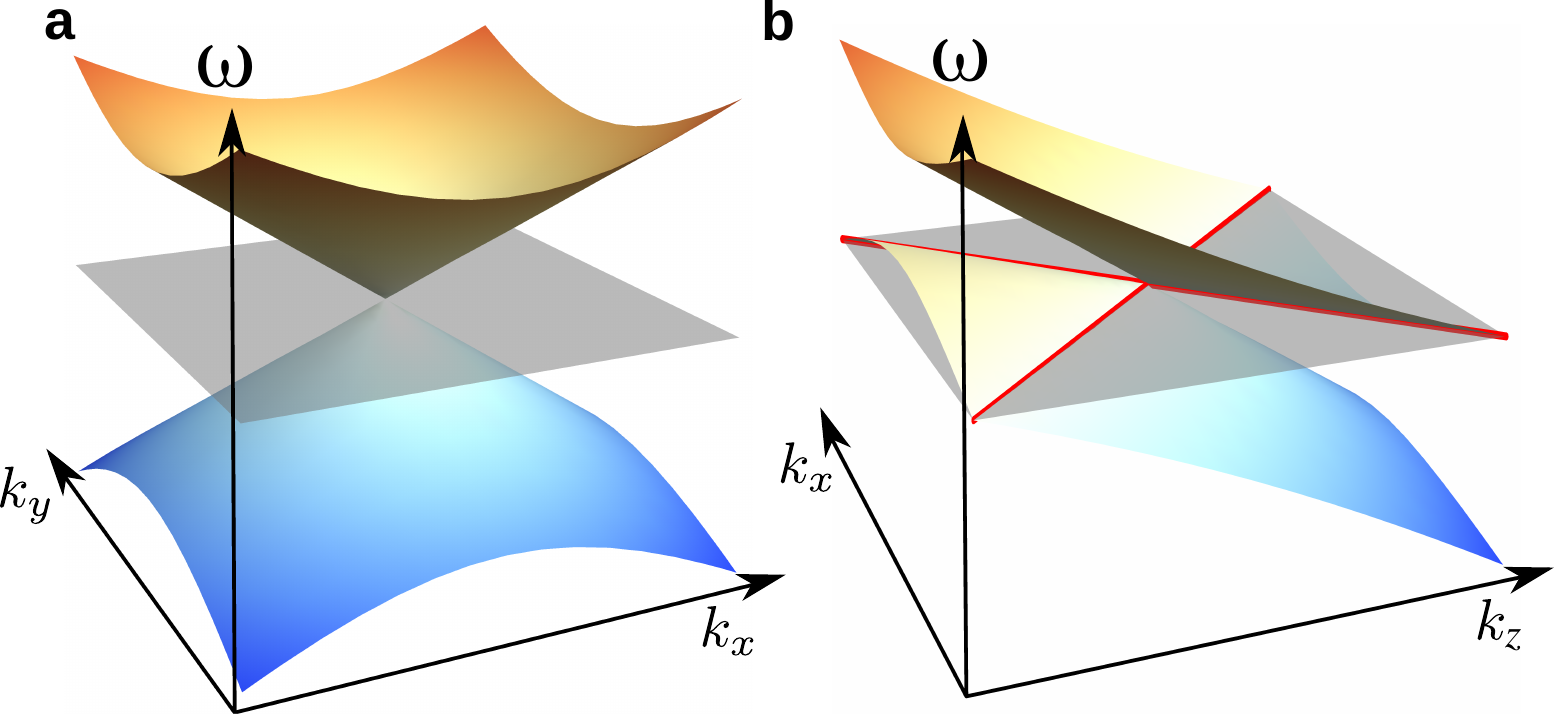}
\caption{Two-dimensional cuts of a Type-II Weyl point. Grey planes indicate the Weyl point frequency. (a) In the transverse plane ($k_z=0$) the two bands touch at a point. (b) Tilted dispersion (tilt parameter $b=1/2$; see Supplementary Material for precise definition) in longitudinal ($k_z$) direction generates a conical isofrequency surface (red lines).}

\end{figure}

\newpage

\begin{figure}
\includegraphics[width=6in]{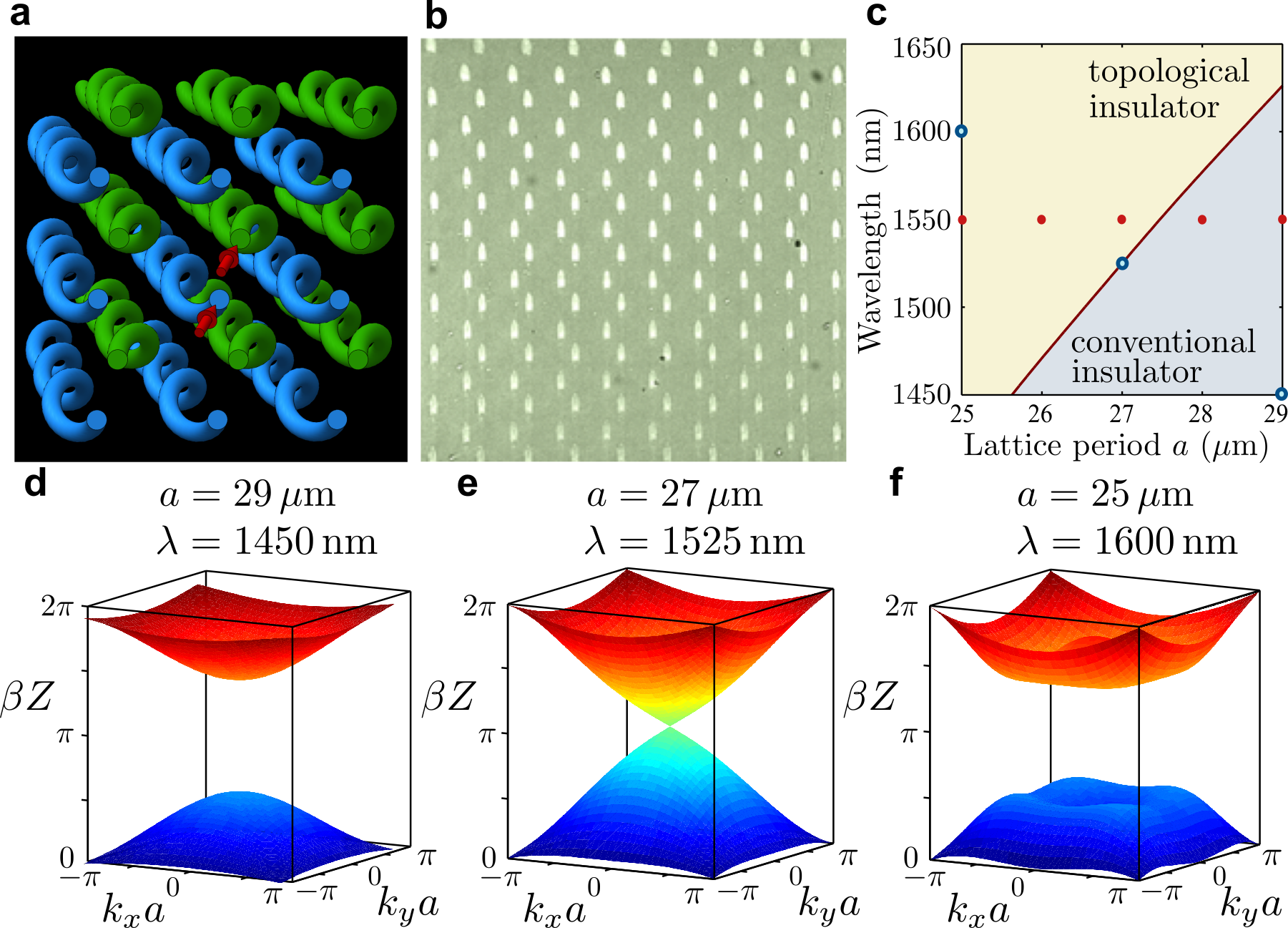}
\caption{Waveguide array structure along with corresponding bulk photonic band structures. (a) Schematic diagram of the waveguide array in three dimensions, with red arrows indicating the waveguides into which light is injected. (b) Microscope image of the output facet of structure.  Note that this only shows a cross-section of the structure; the helicity is not observable here. (c) topological phase diagram of the lattice versus the lattice period $a$ and operating wavelength $\lambda$; (d-f) Bulk band structures (isofrequency surfaces) for the topologically trivial case, the transition (Weyl) point, and the topological case, at $a=29, 27, 25\micron$, at wavelengths $1450\nm, 1525\nm$ and $1600\nm$, respectively.  The open circles in the phase diagram shown in (c) correspond to the band structures in (d-f)}
\end{figure}

\newpage

\begin{figure}
\includegraphics[width=6in]{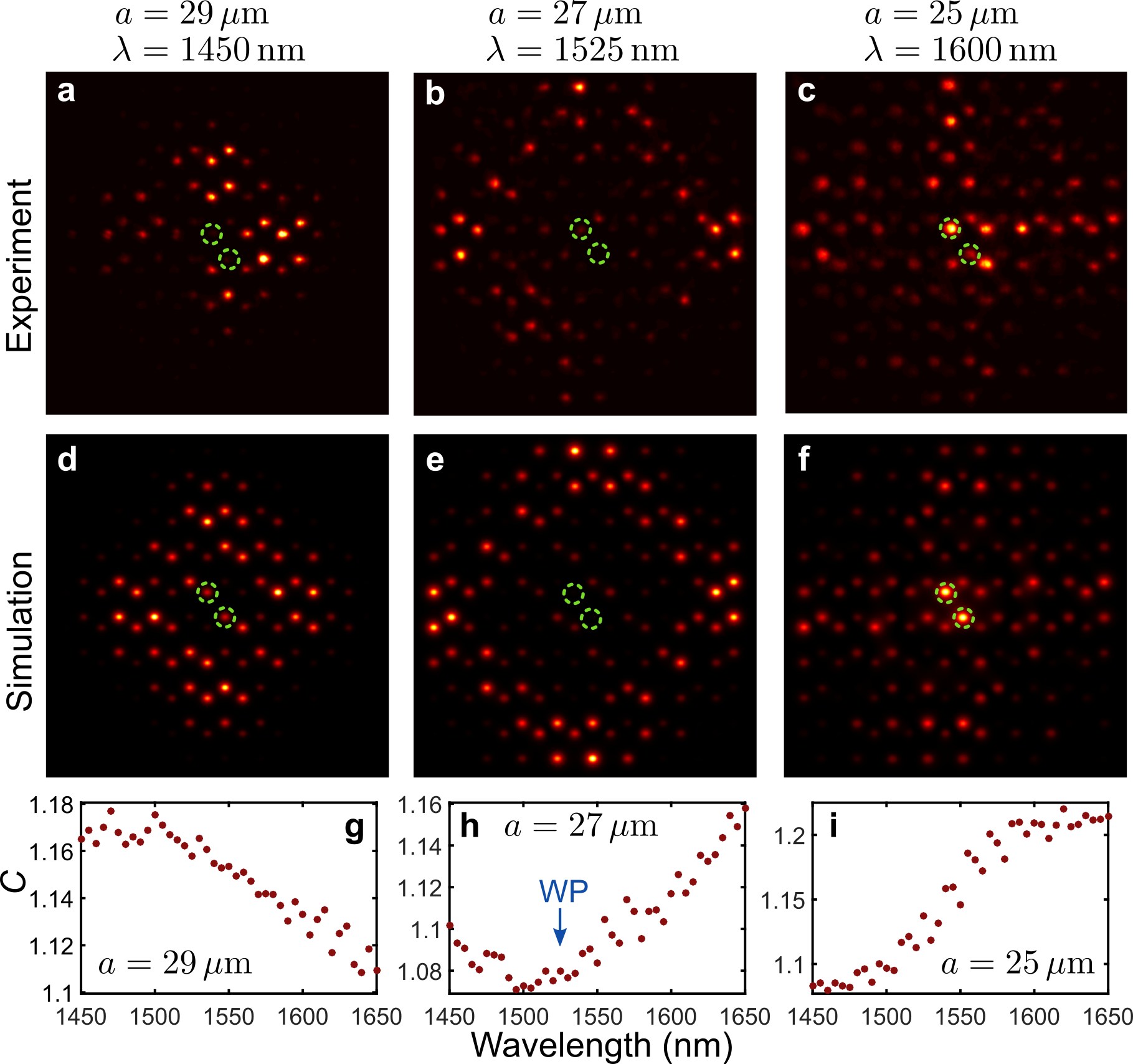}
\caption{Observation of conical diffraction resulting from the presence of the type-II Weyl point.  (First row: a-c) Output optical wavefunction in the topologically trivial regime, at the Weyl point, and in the topological regime for $a=29, 27, 25\micron$, at wavelengths $1450\nm, 1525\nm$ and $1600\nm$, respectively.  The green circles indicate the position of the input waveguides.  Clear conical diffraction is observed in (b), at the Weyl point.  (Second row: d-f)  Full-wave simulations corresponding to the parameters of (a-c).  (Third row: g-i) Plot of the quantity $C(\lambda)$, which quantifies how ring-like the wavefunction for $a=29, 27, 25\micron$, as a function of wavelength from $1450$--$1650\nm$.  In (d) and (f), there is no clear minimum, indicating the lack of a Weyl point within this wavelength range.  However, $(d)$ shows a clear minimum - at the wavelength where the Weyl point lies.  This minimum corresponds to the wavefunction shown in (b).
}
\end{figure}

\newpage

\begin{figure}
\includegraphics[width=6in]{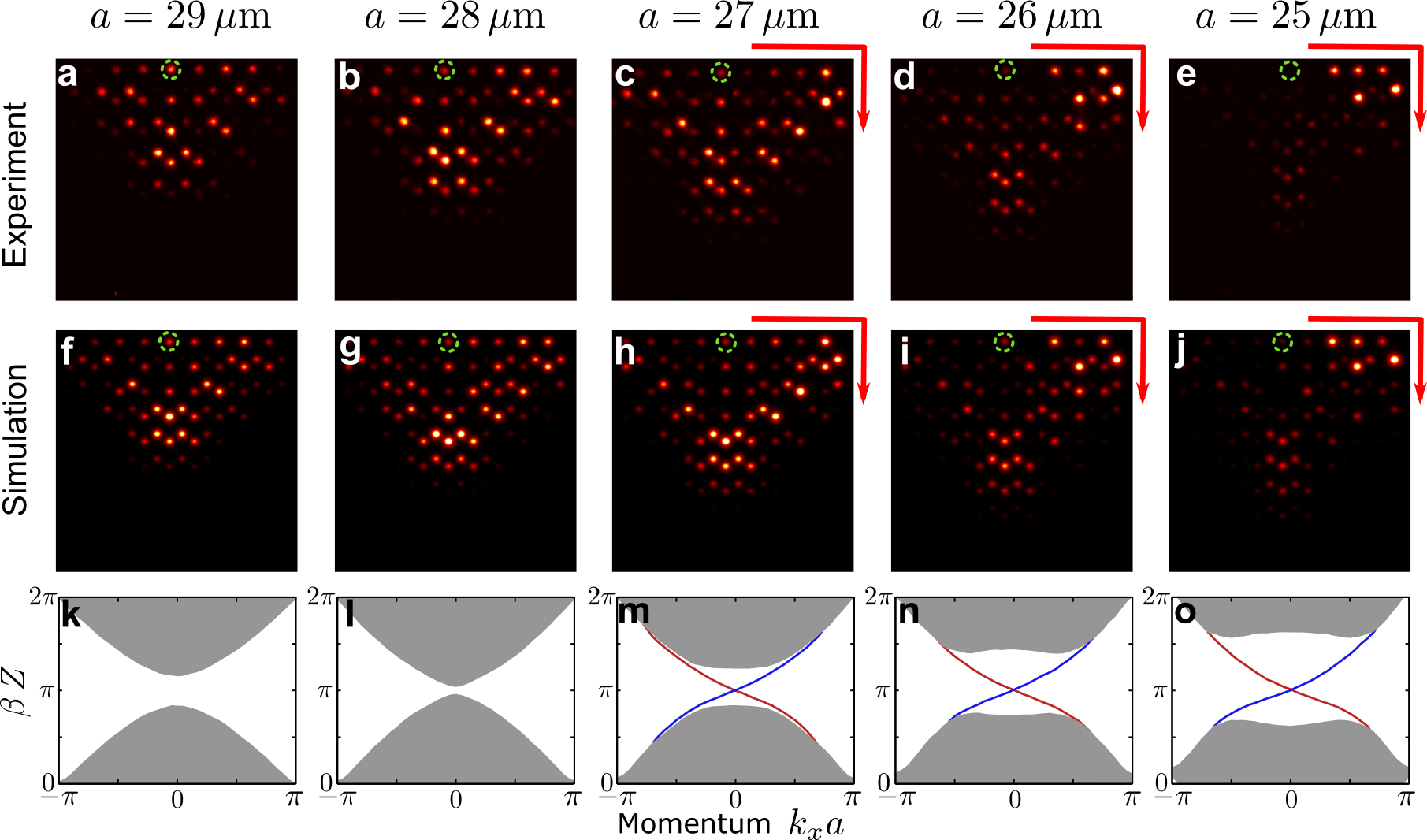}
\caption{Demonstration of the topological transition associated with the Weyl point.  The green circles indicate the position of the input waveguides.  (First row: a-e) Resulting output wavefunction when light is input at the center of the top edge of the structure at wavelength $1550\nm$, for decreasing $a=29, 28, 27, 26, 25\micron$.  Figure (a) lies deep in the topologically trivial regime, where light that is input at the center of the edge simply spreads into the bulk; whereas (e) lies deep in the topological regime, and light stays confined to the edge and moves in a clockwise sense.  Corresponding full-wave beam-propagation simulations are shown in (second row: f-j) and show strong agreement with experimental results.  (Third row: h-o) show corresponding band structures, indicating the emergence of the topological gap, and therefore the chiral edge state (i.e., Fermi arc surface state), as a function decreasing lattice constant $a$.  The red and blue curves indicate edge states on the top and bottom of the sample, respectively.  The trajectory of the edge state wavepackets along the edge are schematically indicated by a red arrow.  }
\end{figure}

\clearpage

\section{Supplementary Information: Mapping the 2D Dirac Hamiltonian to a 3D Type-2 Weyl Hamiltonian}

As described in the main text, the steady-state diffraction of light
through the waveguide array, at a given operating frequency $\omega$,
can be described by a paraxial approximation.  Given the field
amplitude $\mathcal{E}(x,y,z)$ with a certain fixed mode polarization
(see the discussion in the main text), we define a slowly-varying
envelope field $\psi(x,y,z)$ by
\begin{equation}
  \mathcal{E}(x,y,z) = \psi(x,y,z) \,e^{ik_0 z},
  \label{envelope}
\end{equation}
where $k_0 = n_0\omega/c$, and $n_0$ is the ambient refractive index
of the medium.  In the limit of negligible backscattering along the
$z$ axis, $\psi(x,y,z)$ satisfies a 2D Schr\"odinger equation
\begin{equation}
  i \frac{\partial\psi}{\partial z}
  \approx \hat{H} \psi(x,y,z), \;\;\mathrm{where}\;\;
  \hat{H} =
  - \frac{1}{2k_0} \nabla_\perp^2 - \frac{k_0}{n_0}\delta n(x,y,z).
  \label{2dschrod}
\end{equation}
Here, $x$ and $y$ are the directions perpendicular to the waveguide
axis, while $z$ plays the role of time.  $\nabla_\perp^2$ is the 2D
Laplacian, and $\delta n = n - n_0$ is the refractive index shift.

For a photonic lattice in which $\delta n$ is periodic in $z$, with
period $L_z$, Eq.~(\ref{2dschrod}) turns into a Floquet problem.  The
Floquet eigenstates satisfy
\begin{align}
  \psi(x,y,z+L_z) &= \psi(x,y,z) \, e^{i\beta z},\\
  \hat{H}_F \psi &= \beta \psi,
\end{align}
where $\hat{H}_F$ is the Floquet Hamiltonian and its eigenvalue
$\beta$ is the Floquet quasi-energy.  The Floquet Hamiltonian is
defined using the $z$-evolution operator over one period (analogous to
the usual time-evolution operator):
\begin{equation}
  \exp(i\hat{H}_FL_z) = \mathcal{T} \exp\left[\,i\int_0^{L_z} dz
    \,\hat{H}(x,y,z)\right].
\end{equation}
Referring back to Eq.~(\ref{envelope}), we see that a 2D Floquet
eigenstate with quasi-energy $\beta$ corresponds to an electromagnetic
Bloch wave of the underlying 3D photonic structure, having Bloch
wave-vector component $k_z$, where
\begin{equation}
  \beta = k_z - k_0.
  \label{kz}
\end{equation}

The specific photonic lattice we are interested in (see the main text)
has a Floquet Hamiltonian $\hat{H}_F$ that can be tuned to a
topological transition at some frequency $\omega_0$.  To lowest order
in the detuning $\delta \omega \equiv \omega - \omega_0$, $\hat{H}_F$
is described by a Dirac equation,
\begin{equation}
  \hat{H}_F \approx
  v_d \big(k_x \hat{\sigma}_x +  k_y \hat{\sigma}_y\big)
  + b\, \frac{n_0 \delta \omega}{c}\, \hat{\sigma}_z,
  \label{Dirac2D}
\end{equation}
for some dimensionless real constants $v_d$ and $b$.  The extra factor
of $n_0/c$ in the last term is for later convenience.  As our
experimental and simulation results show, the system can be tuned
across the transition point---i.e., between negative and positive
Dirac mass---by varying $\omega$.

The Floquet eigenproblem of Eq.~(\ref{Dirac2D}) depends on the
detuning $\delta \omega$ as an implicit parameter, and yields the
Floquet quasi-energy $\beta$ as an eigenvalue.  To make contact with
the Weyl Hamiltonian, we must re-arrange it into an equation with
$\delta \omega$ as the eigenvalue and $\beta$ as a parameter.  First,
we must re-parameterize Eq.~(\ref{kz}) in terms of the frequency
detuning:
\begin{equation}
  \beta = \delta k_z - \frac{\delta\omega}{c_0}, \;\;\;\mathrm{where}\;\;
  \left\{\begin{array}{rl}
  c_0 &\equiv \displaystyle c/n_0, \\
  \delta k_z &\equiv \displaystyle k_z - \omega_0/c_0.
\end{array}\right.
  \label{beta2}
\end{equation}
Using this, we re-arrange the Floquet eigenproblem of
Eq.~(\ref{Dirac2D}) to obtain
\begin{equation}
  \Big[v_d\, (k_x \hat{\sigma}_x +  k_y \hat{\sigma}_y) - \delta k_z \hat{I} \Big] \, \psi = - ( \hat{I} + b\, \hat{\sigma}_z)\, \frac{\delta \omega}{c_0}\, \psi,
\end{equation}
where $\hat{I}$ denotes the identity matrix.  This has the form of a
generalized eigenvalue problem.  To convert it into a Hamiltonian
eigenvalue problem, we seek to factorize the operator on the
right-hand side and rescale the state vectors:
\begin{equation}
  \varphi = \hat{\mathcal{W}} \,\psi, \;\;\;
  \hat{\mathcal{W}}^2 = -c_0^{-1} (\hat{I} - b\, \hat{\sigma}_z).
  \label{W}
\end{equation}
This would then satisfy
\begin{align}
  \hat{H}' \varphi &= \delta \omega\, \varphi,\\
  \hat{H}' &\equiv \hat{\mathcal{W}}^{-1}
  \Big[v_d\, (k_x \hat{\sigma}_x +  k_y \hat{\sigma}_y)
    - \delta k_z \hat{I} \Big]
  \hat{\mathcal{W}}^{-1}.
  \label{hprime}
\end{align}
Assuming that $|b| < 1$ (see below for more discussion), we can
directly verify that appropriate re-scaling operators, consistent with
Eq.~(\ref{W}), are
\begin{align}
  \hat{\mathcal{W}} &=
  i\left(\frac{1}{2c_0} - \sqrt{\frac{1-b^2}{4c_0^2}}\right)^{1/2} \hat{I}
    + i\left(\frac{1}{2c_0} + \sqrt{\frac{1-b^2}{4c_0^2}}\right)^{1/2}
    \hat{\sigma}_z, \\
    \hat{\mathcal{W}}^{-1} &=
    \frac{c_0}{\sqrt{1-b^2}}\,
    \left[
      i\left(\frac{1}{2c_0} - \sqrt{\frac{1-b^2}{4c_0^2}}\right)^{1/2}
      \hat{I}
        - i\left(\frac{1}{2c_0} + \sqrt{\frac{1-b^2}{4c_0^2}}\right)^{1/2}
    \hat{\sigma}_z \right].
    \label{winv}
\end{align}
Due to the assumption that $|b| < 1$, all terms in square roots are
all positive.  Using Eq.~(\ref{winv}) on Eq.~(\ref{hprime}) yields
\begin{equation}
  \hat{H}' = \frac{c_0}{\sqrt{1 -b^2}}
  \Big[v_d\, (k_x \hat{\sigma}_x +  k_y \hat{\sigma}_y)\Big]
  + \frac{c_0^2}{1-b^2} \, \delta k_z\,
  \left(\gamma \hat{I} - |b| \hat{\sigma}_z\right).
\end{equation}
We simplify this by defining
\begin{align}
  v_d' &= \frac{v_d c_0}{\sqrt{1-b^2}} \\
  k_z' &= -\frac{|b|/v_d}{\sqrt{1-b^2}}\, \delta k_z
\end{align}
Thus, the Hamiltonian finally reduces to
\begin{equation}
  \hat{H}' =
  v_d'\, \left(k_x \hat{\sigma}_x +  k_y \hat{\sigma}_y
  + k_z' \hat{\sigma}_z - \frac{k_z'}{|b|} \hat{I} \right). \label{final_dirac}
\end{equation}
This is a type-II Weyl Hamiltonian.  Lorentz invariance is broken by
the $-(k_z'/|b|)\, \hat{I}$ term (which cannot be eliminated without
violating our earlier assumption that $|b| < 1$).

We can perform an order-of-magnitude estimate for the dimensionless
Lorentz invariance breaking parameter $b$. In the $a=27\mu$m sample the Dirac point occurs at $\lambda_1 \approx 1525$nm. Detuning to $\lambda_2 \approx 1550$nm opens up a band gap of size $\approx \frac{\pi}{2Z}$, where $Z=1$cm is the helix pitch. Therefore $\Delta \omega n_0 b / c \approx \frac{\pi}{4Z}$. Using $k_0 = 2 \pi n_0 \omega / c = 2 \pi / \lambda$ and $\omega = c/(n_0 \lambda)$, $\Delta \omega = \frac{c}{n_0}(\frac{1}{\lambda_1}-\frac{1}{\lambda_2}) \approx \frac{c}{n_0 \lambda} \frac{\delta \lambda}{\lambda}$ and $b = \frac{\pi \lambda}{4Z} \frac{\lambda}{\delta \lambda} \approx 7\times10^{-3} \ll 1$ so we are well within validity of assumption $|b|<1$. Note that the scale of $b$ is set by $\delta n / n_0 \sim 10^{-3}$; in the ``photonic crystal'' limit when propagation in no longer paraxial and backscattering cannot be ignored, the constraint $|b|<1$ may be violated. 

\section{Edge states and Fermi arcs}

Here we relate the observation of a topological transition and edge states in the main text to the existence of Fermi arcs in the type-II Weyl Hamiltonian. Consider a semi-infinite lattice with surface normal in the $y$ direction. Looking for modes localized in the $y$ direction, one can define a surface Bloch Hamiltonian $\hat{H}_{2D}(k_x,\omega)$ with eigenvalues $k_z$, and using the same procedure outlined above obtain an equivalent eigenvalue problem for $\omega$. The Fermi arcs will form surfaces in the $(k_x,k_z,\omega)$ parameter space bounded by the bulk band edges. 

To identify Weyl points based on the surface spectrum, instead of considering these full eigenvalue surfaces in 3D space it is more convenient to consider the surface states at a fixed frequency $\omega$, which form lines (isofrequency contours) within the 2D Brillouin zone. As long as no bulk states are encountered, contours corresponding to topologically trivial surface states must be closed, while Fermi arcs form open contours terminating at Weyl points of opposite chirality~\cite{fermi_arc_identification,fermi_arc_identification_2}. A sufficient criterion to identify a type-I Weyl point is to consider any closed loop within the Brillouin zone that avoids the bulk bands and count the number of intersecting isofrequency contours: if the number of intersecting lines is odd, there is a Fermi arc terminating at a Weyl point encircled by the loop.

Because of the tilted dispersion at a type-II Weyl point, an encircling loop at a single frequency cannot avoid bulk states and this procedure has to be modified according to Ref.~\cite{weyl2_fermi_arc}: instead of a loop at a single frequency, one should instead divide the loop into two segments within the bulk band gap at frequencies on either side of the Weyl point, e.g. as in Figs. 4(k,o) of the main text. When $\delta \omega > 0$ (Fig. 4k) there are no surface states that can cross the segment. When $\delta \omega < 0$ (Fig. 4o) a single surface state (per edge) will cross the segment if the projection of the Weyl point onto the ($k_x,k_z \sim \beta Z$) plane is encircled. Thus, the odd number of crossings induced by the topological transition of the 2D Floquet Hamiltonian signifies the type-II Weyl point.

%\begin{thebibliography}{99}

%\bibitem{fermi_arc_identification}
%B. Q. Lv et al., ``Experimental Discovery of Weyl semimetal TaAs,'' Phys. Rev. X 5, 031013 (2015).

%\bibitem{fermi_arc_identification_2}
%I. Belopolski et al., ``Criteria for directly detecting topological Fermi arcs in Weyl semimetals,'' Phys. Rev. Lett. {\bf 116}, 066802 (2016).

%\bibitem{weyl2_fermi_arc}
%I. Belopolski et al., ``Fermi arc electronic structure and Chern numbers in the type-II Weyl semimetal candidate Mo$_x$W$_{1-x}$Te$_2$,'' Phys. Rev. B {\bf 94}, 085127 (2016).

%\end{thebibliography}

\clearpage

\end{document}